%% file: two_loop.tex
\documentclass[a4paper,12pt]{article}

\usepackage{xspace}

\usepackage[T1]{fontenc}
\usepackage[utf8]{inputenc}

\usepackage[ngerman, english]{babel}

\usepackage{xcolor}
\usepackage{graphicx}

\usepackage{amsmath}
\usepackage[separate-uncertainty]{siunitx}
\usepackage{slashed}

\usepackage[pdfborderstyle={/S/U/W 1}]{hyperref}

\usepackage{csquotes}
\usepackage[natbib=true, backend=biber, sorting=none, sortcites=true, style=numeric-comp]{biblatex}
\usepackage{glossaries}
\glsdisablehyper

\input{glossary}

\begin{document}

\begin{titlepage}

\begin{flushright}
 DO-TH 17/06\\
 QFET-2017-14\\
 \today
\end{flushright}
\vskip 2.4cm

\begin{center}
\boldmath
{\Large\bf Two loop virtual corrections to $b\to(d,s)\ell^+\ell^-$ and $c\to u\ell^+\ell^-$ for arbitrary momentum transfer}
\unboldmath
\vskip 2.2cm
{\sc Stefan de Boer$^a$}
\vskip .5cm

\vspace{0.7cm}
{\sl ${}^a$Fakult\"at f\"ur Physik,\\
TU Dortmund, Otto-Hahn-Str.4, 44221 Dortmund, Germany\\
}
\vspace{3\baselineskip}

\vspace*{0.5cm}

\end{center}

\begin{abstract}

Non-factorizable two loop corrections to heavy to light flavor changing neutral current transitions due to matrix elements of current-current operators are calculated analytically for arbitrary momentum transfer.
This extends previous works on \mbox{$b\to(d,s)\ell^+\ell^-$} transitions.
New results for \mbox{$c\to u\ell^+\ell^-$} transitions are presented.
Recent works on polylogarithms are used for the master integrals.
For \mbox{$b\to s\ell^+\ell^-$} transitions, the corrections are most significant for the imaginary parts of the effective Wilson coefficients in the large hadronic recoil range.
Analytical results and ready-to-use fitted results for a specific set of parameters are provided.

\end{abstract}

\end{titlepage}

\section{Introduction}

Recently, several discrepancies with the \gls{sm} have been revealed in \mbox{$b\to s\ell^+\ell^-$} induced decays, e.g., \cite{Blake:2017wjz}.
However, those that do not involve lepton flavor universality violation are not definitely settled due to poorly controlled non-perturbative effects of \gls{qcd}.
Exemplary, non-factorizable corrections to form factors are commonly considered as one unreducible uncertainty while interpreting \gls{bsm} physics.
Furthermore, in \mbox{$c\to u\ell^+\ell^-$} transitions, non-factorizable corrections yield the largest perturbative contribution due to the \gls{gim} cancellation of factorizable contributions \cite{Greub:1996wn,deBoer:2015boa,Feldmann:2017izn}.
In this paper, we improve on the current state by calculating the two loop virtual corrections to any heavy to light transition, including $c$ and $b$ decays, induced by current-current operators for arbitrary invariant dilepton mass $q^2$.

Partonic transitions are a first approximation to the corresponding inclusive hadronic decays in the framework of an \gls{ope}.
This approximation is applicable away from resonance regions and up to power corrections.
The resonance regions can be handled by appropriate kinematical cuts and power corrections can be treated within a heavy quark expansion (an expansion in inverse powers of the heavy quark mass, see \cite{Ghinculov:2003qd} for \mbox{$b\to s$} transitions).
Furthermore, perturbative results are the basis for estimates of non-perturbative effects in inclusive and exclusive hadronic decays, e.g., \cite{Benzke:2010js,Khodjamirian:2010vf}.

Several calculations were performed for \mbox{$b\to(d,s)$} transitions:
In \cite{Asatryan:2001zw} and \cite{Asatrian:2003vq} \mbox{$b\to s$} and \mbox{$b\to d$} transitions, respectively, were computed for small $q^2$.
The calculation of \mbox{$b\to s$} transitions for large $q^2$ was accomplished in \cite{Greub:2008cy}.
A seminumerical approach was employed in \cite{Ghinculov:2003qd} to present results based on \mbox{$b\to s$} transitions for the full $q^2$ range.
In \cite{Seidel:2004jh} \mbox{$b\to d$} transitions for any $q^2$ range were computed, extending \cite{Asatrian:2003vq}.
For \mbox{$c\to u$} transitions, see \cite{deBoer}.
We emphasize that available results for \mbox{$b\to(d,s)$} transitions only cover different limits, i.e.\ zero masses, small and large $q^2$ ranges, and that results for \mbox{$c\to u$} transitions have become available only recently \cite{deBoer,Feldmann:2017izn}.
The analytic calculation presented in this paper covers the full $q^2$ range, arbitrary masses and electric charges.

Generally, the effective weak Lagrangian for heavy to light quark \gls{fcnc} transitions \mbox{$h\to l$} at the low-energy scale $\mu$ is written as
\begin{align}\label{eq:L_eff_mu}
 \mathcal L_\text{eff}=\frac{4G_F}{\sqrt 2}\sum_q\lambda_q\left(C_1P_1^{(q)}+C_2P_2^{(q)}+\sum_{i=3}^{10}C_iP_i\right)\,,
\end{align}
where the sum is over light quark fields $q$ with masses below $\mu$, and products of \gls{ckm} matrix elements $\lambda_q=V_{q(d,s)}^*V_{qb},V_{cq}^*V_{uq}$ for \mbox{$b\to(d,s)$} and \mbox{$c\to u$} transitions, respectively.
Here, $C_i$ are the Wilson coefficients and the physical operators $P_i$, which are relevant for this paper, read
\begin{align}
 &P_1^{(q)}=(\bar l_L\gamma_{\mu_1}T^aq_L)(\overline q_L\gamma^{\mu_1}T^ah_L)\,,\label{eq:P1}\\
 &P_2^{(q)}=(\bar l_L\gamma_{\mu_1}q_L)(\overline q_L\gamma^{\mu_1}h_L)\,,\label{eq:P2}\\
 &P_7=\frac e{g_s^2}m_h(\bar l_L\sigma^{\mu_1\mu_2}h_R)F_{\mu_1\mu_2}\,,\label{eq:P7}\\
 &P_9=\frac{e^2}{g_s^2}(\bar l_L\gamma_{\mu_1}h_L)(\overline\ell\gamma^{\mu_1}\ell)\,,\label{eq:P9}
\end{align}
where $q_{L/R}=\tfrac12(1\mp\gamma_5)q$, $\sigma^{\mu_1\mu_2}=\tfrac i2[\gamma^{\mu_1},\gamma^{\mu_2}]$, and $T^a$ are the $SU(3)_C$ generators normalized to $\mathrm{Tr}[T^aT^b]=\tfrac{\delta^{ab}}2$.
Furthermore, $G_F$ is the Fermi constant, $F_{\mu_1\mu_2}$ denotes the electromagnetic field strength tensor, and $g_s$ and $e$ are the strong and electromagnetic coupling constants, respectively.

We calculate the two loop \gls{qcd} matrix elements of $P_{1/2}$, in terms of form factors (i.e., for inclusive decays, effective Wilson coefficients) multiplying the matrix elements of $P_{7,9}$, for $h\to l\ell^+\ell^-$ transitions.
The result is valid for a general class of heavy to light transitions with arbitrary invariant momentum transfer and masses, when the mass of the light quark is neglected.
This includes the transitions \mbox{$b\to(d,s)$} via $(u\bar u,c\bar c)$ loops, and \mbox{$c\to u$} via $(d\bar d,s\bar s)$ loops, where the loop quark-antiquark pair $(q\bar q)$ is annihilated and a photon is emitted, which may then couple to a lepton pair.
Note that the computation of two loop matrix elements presented in this paper is not restricted to \gls{sm} applications, see \cite{deBoer} for an example in leptoquark models.
We use the recent works \cite{Bell:2014zya} and \cite{Frellesvig:2016ske} for the \glspl{mi} and their numerical evaluation, respectively.

We outline our calculation in the section~\ref{sec:outline_calculation}, see also \cite{deBoer}.
The numerical evaluation is detailed in section~\ref{sec:numerical_evaluation}.
Results are given in section~\ref{sec:results}, where we also comment on the phenomenological impact for \mbox{$b\to(d,s)$} transitions.
For the phenomenology of \mbox{$c\to u$} transitions, see \cite{deBoer}.
Appendix~\ref{app:supplemented_files} contains a description of supplemented files, which encode the results of this paper.

\section{Outline of the calculation}
\label{sec:outline_calculation}

\begin{figure}
 \centering
 \includegraphics[trim=4cm 22cm 6cm 3cm,clip,width=0.9\textwidth]{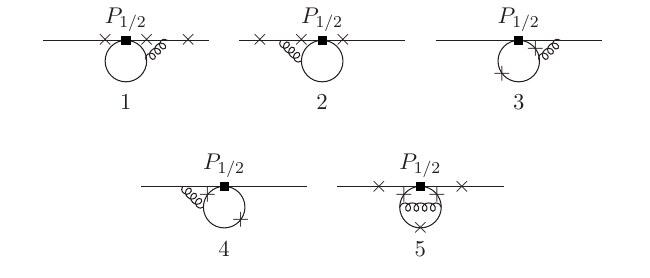}
 \caption{Diagrams for heavy to light quark transitions at two loop \gls{qcd}.
 The boxes denote operator insertions of $P_{1/2}$.
 The crosses indicate the emission of a photon, which may then couple to a lepton pair.}
 \label{fig:diagrams1to5}
\end{figure}

\begin{figure}
 \centering
 \includegraphics[trim=4cm 24cm 6cm 2cm,clip,width=0.9\textwidth]{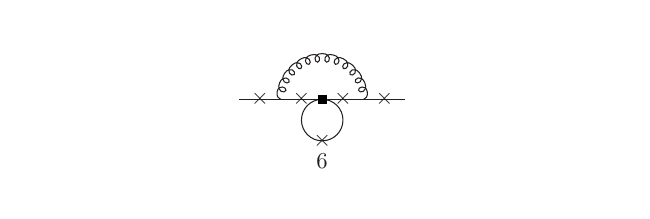}
 \caption{A sixth class of diagrams, see figure~\ref{fig:diagrams1to5} and text.}
 \label{fig:diagram6}
\end{figure}

In this section, we outline the calculation of the diagrams shown in figure~\ref{fig:diagrams1to5}.
Each of the five subsets represents a gauge invariant class.
A sixth class, shown in figure~\ref{fig:diagram6}, preserves the operator structure of $P_9$, hence it is commonly considered as a correction to the matrix element of $P_9$ \cite{Asatryan:2001zw} and not included in the calculation presented in this paper.
Nevertheless, we have checked that in this class only the diagram with a photon coupling to the loop of the quark-antiquark pair is non-zero and factorizes into two one loop integrals.
It is the only diagram with infrared and collinear singularities.

We calculate the diagrams in figure~\ref{fig:diagrams1to5} with insertions of $P_2$.
Insertions of $P_1$ are then given by color factors due to additional generators in the definition of the operator:
The expressions for the first four and last (two) classes are multiplied with $\tfrac{-1}6$ and $\tfrac43$, respectively.

The matrix element for an off-shell photon $\gamma^*$ can be decomposed as \cite{Seidel:2004jh}
\begin{align}
 \langle\gamma^*(q,\mu)l(p_l)|P_2|h(p_h)\rangle&=\langle l(p_l)|\bar lX^\mu h|h(p_h)\rangle\nonumber\\
 &=\left(\frac{\alpha_s}{4\pi}\right)^2F^{(q)}(q^2)\langle l(p_l)|\bar l_Lq^\mu h_L|h(p_h)\rangle\nonumber\\
 &+\left(\frac{\alpha_s}{4\pi}\right)^2F^{(7)}(q^2)\langle l(p_l)|\bar l_L\sigma^{\mu\nu}q_\nu h_R|h(p_h)\rangle\nonumber\\
 &+\left(\frac{\alpha_s}{4\pi}\right)^2F^{(9)}(q^2)\langle l(p_l)|\bar l_L(q^2\gamma^\mu-q^\mu\slashed q)h_L|h(p_h)\rangle\,,
\end{align}
where $q^\mu=(p_h-p_l)^\mu$ is the transferred momentum, and $X^\mu$ is the sum of the amputated diagrams in figure~\ref{fig:diagrams1to5}.

The scalar form factors $F^{(i)}$ are given as \cite{Seidel:2004jh}
\begin{align}
 F^{(i)}(q^2)=\mathrm{Tr}[P_i^\mu X_\mu]
\end{align}
with the projectors
\begin{align}
 P_i^\mu=(\slashed p_h+m_h)(C_{i1}q^\mu+C_{i2}p_h^\mu+C_{i3}\gamma^\mu)\slashed p_l
\end{align}
and coefficients $C_{ij}$.
The factors $(\slashed p_h+m_h)$ and $\slashed p_l$ reflect the on-shell conditions for the external quarks.
The form factors for $\langle l\ell\ell|P_2|h\rangle$, thus the effective Wilson coefficients, are proportional to $F^{(7,9)}$.
By gauge invariance, the form factor $F^{(q)}$ vanishes for each class, which we have checked.

For the calculation, we utilize the computer programs \texttt{FORM~4.0} \cite{Kuipers:2012rf} and \texttt{REDUZE~2} \cite{vonManteuffel:2012np}.
We use the former for algebraic manipulations, e.g.\ the tensor algebra, and the latter is used for the reduction to \glspl{mi}.
The program \texttt{REDUZE~2} is based on the Laporta algorithm \cite{Laporta:2001dd} which employs \gls{ibp} identities \cite{Chetyrkin:1981qh} and \gls{li} identities \cite{Gehrmann:1999as}.
Indeed, \gls{li} identities can be written as linear combinations  of \gls{ibp} identities \cite{Lee:2008tj}.
We have calculated relations for each diagram based on \gls{li} identities by hand and checked them against the reduction tables built with \texttt{REDUZE~2}.

We find that the following diagrams in figure~\ref{fig:diagrams1to5} vanish:
In the fifth class, both diagrams with photons attached to the external lines vanish.
The diagram with a photon emitted from the heavy quark line in the first class is zero.
For $F^{(7)}$, all diagrams with photons attached to the external lines vanish.
Furthermore, all diagrams of the fifth class vanish for $F^{(7)}$.
This implies that the $\langle P_1\rangle$ induced dipole form factor is given by $\tfrac{-1}6F^{(7)}$.

As for the set of the \glspl{mi}, we match a subset of the integrals calculated in \cite{Bell:2014zya}.
A canonical set is given in \cite{Greub:2008cy}, where the \glspl{mi} are calculated for large $q^2$, yet the set of integrals is not minimal.
While matching the set of \cite{Greub:2008cy} onto the \glspl{mi} taken from \cite{Bell:2014zya}, we find additional relations among the former, e.g.\ for the integral $I_{d23}$ and the one of equation (A12) in \cite{Greub:2008cy}.
Furthermore, we do not encounter the integrals of equations (A5), (A8), and (A14) in \cite{Greub:2008cy}.

With the \glspl{mi} from \cite{Bell:2014zya} the unrenormalized form factors are expressed in terms of (generalized) \glspl{hpl}, see the next section for the numerical evaluation.
The prescription for the renormalization is described in, e.g., \cite{Asatryan:2001zw}.
Accordingly, the operator renormalization constants are written as
\begin{align}
 &Z_{ij}=\delta_{ij}+\delta Z_{ij}\,,\nonumber\\
 &\delta Z_{ij}=\frac{\alpha_s}{4\pi}\left(a_{ij}^{01}+\frac1{\epsilon}a_{ij}^{11}\right)+\frac{\alpha_s^2}{(4\pi)^2}\left(a_{ij}^{02}+\frac1{\epsilon}a_{ij}^{12}+\frac1{\epsilon^2}a_{ij}^{22}\right)+\mathcal O(\alpha_s^3)\,,
\end{align}
where the dimension is $4-2\epsilon$.
Extending the set of the physical operators $P_i$ by the evanescent operators $E_{11,12}$ defined as
\begin{align}
 &E_{11}=\left(\bar l_L\gamma_{\mu_1}\gamma_{\mu_2}\gamma_{\mu_3}T^aq_L\right)\left(\bar q_L\gamma^{\mu_1}\gamma^{\mu_2}\gamma^{\mu_3}T^ah_L\right)-16P_1\,,\\
 &E_{12}=\left(\bar l_L\gamma_{\mu_1}\gamma_{\mu_2}\gamma_{\mu_3}q_L\right)\left(\bar q_L\gamma^{\mu_1}\gamma^{\mu_2}\gamma^{\mu_3}h_L\right)-16P_2\,.
\end{align}
the coefficients $a_{ij}$ are compactly written as
{\renewcommand{\arraystretch}{1.5}\setcounter{MaxMatrixCols}{12}
\begin{align}\label{eq:a11}
 a^{11}|_{i=1,2}=
 \begin{pmatrix}
  -2  &  \frac43  &  0  &  -\frac19  &  0  &  0  &  0  &  0  &  -\frac89q_i  &  0  &  \frac5{12}  &  \frac29 \\
  6   &  0        &  0  &  \frac23   &  0  &  0  &  0  &  0  &  -\frac23q_i  &  0  &  1           &  0 \\
 \end{pmatrix}\,,
\end{align}
}%
and
\begin{align}
 &-6\,a_{17}^{12}=a_{27}^{12}=2q_i-\frac8{27}q_e\,,\quad a_{19}^{12}=-\frac29q_i-\frac{44}{243}q_e\,,\quad a_{29}^{12}=\frac{16}3q_i+\frac{88}{81}q_e\,,\label{eq:a12}\\
 &a_{19}^{22}=\frac{92}9q_i-\frac{16}{27}n_f\,q_i+\frac8{81}q_e\,,\quad a_{29}^{22}=\frac{14}3q_i-\frac49n_f\,q_i-\frac{16}{27}q_e\,.
\end{align}
Here, $P_4=(\bar l_L\gamma_{\mu_1}T^ah_L)\sum_{\{q:m_q<\mu\}}(\overline q\gamma^{\mu_1}T^aq)$, $n_f$ is the number of flavors and $q_{e,i}$ are the charges of the external and internal quarks, respectively, i.e.\ $q_e=\tfrac{-1}3$, $q_i=\tfrac23$ for \mbox{$b\to(d,s)$} and $q_e=\tfrac23$, $q_i=\tfrac{-1}3$ for \mbox{$c\to u$} transitions.
The coefficients $a_{ij}^{11,22}$ can be obtained from the \gls{lo} \gls{adm} $\gamma^{(0)}$ and the coefficients $a_{ij}^{12}$ from \cite{Chetyrkin:1997gb},
\begin{align}
 &a^{11}=\frac12\gamma^{(0)}\,,&&a^{12}=\frac14\gamma^{(1)}+\frac12\hat b\cdot\hat c\,,&&a^{22}=\frac18\gamma^{(0)}\cdot\gamma^{(0)}-\frac14\beta_0\gamma^{(0)}\,,
\end{align}
where $\gamma^{(1)}$ is the \gls{nlo} \gls{adm}, and the mixing via the evanescent operators is found as
{\renewcommand{\arraystretch}{1.5}
\begin{align}
  &\hat b|_{i=1,2}=
 \begin{pmatrix}
  \frac5{12}  &  \frac29 \\
  1           &  0
 \end{pmatrix}\,,&&
  \hat c|_{j=9}=
 \begin{pmatrix}
  \frac{32}9q_i \\
  \frac83q_i
 \end{pmatrix}\,.
\end{align}
}%

The counterterm form factors $F_{i\to(7,9)}^{\text{ct}(7,9)}$, due to the mixing of $P_{1/2}$ into $P_{7,9}$, and the one loop renormalization of $g_s$ in the definition of $P_9$, are \cite{Asatryan:2001zw,Asatrian:2003vq}
\begin{align}
 F_{i\to7}^{\text{ct}(7)}=-\frac{a_{i7}^{12}}{\epsilon}\,,&&F_{i\to9}^{\text{ct}(9)}=-\left(\frac{a_{i9}^{22}}{\epsilon^2}+\frac{a_{i9}^{12}}{\epsilon}\right)-\frac{a_{i9}^{11}\beta_0}{\epsilon^2}\,,
\end{align}
where $\beta_0=11-\tfrac23n_f$.

The counterterms $F_{i\to4\text{quark}}^{\text{ct}(7,9)}$, due to the mixing of $P_{1/2}$ into four-quark operators $P_j$, are defined by \cite{Asatryan:2001zw}
\begin{align}\label{eq:Fct9ito4quark}
 \sum_j\frac{\alpha_s}{4\pi}\frac1{\epsilon}a_{ij}^{11}\langle l\ell\ell|P_j|h\rangle_{1\text{loop}}=-\left(\frac{\alpha_s}{4\pi}\right)^2\left(F_{i\to4\text{quark}}^{\text{ct}(7)}\langle P_7\rangle_\text{tree}+F_{i\to4\text{quark}}^{\text{ct}(9)}\langle P_9\rangle_\text{tree}\right)\,.
\end{align}
We calculate them to $\mathcal O(\epsilon)$ and for insertions of $P_{1,2,4}$, $E_{11,12}$ according to eqs.~(\ref{eq:Fct9ito4quark}) and (\ref{eq:a11}), respectively.
We write
\begin{align}\label{eq:l1}
 I(m_q^2)&=\left(\frac{\mu^2}{m_q^2}\right)^\epsilon\int_0^1\mathrm dz(1-z)z\bigg(-\frac6{\epsilon}+6-6\ln\frac1{1-z(1-z)q^2/m_q^2}-\frac{\pi^2}2\epsilon\nonumber\\
 &+6\epsilon\ln\frac1{1-z(1-z)q^2/m_q^2}-3\epsilon\ln^2\frac1{1-z(1-z)q^2/m_q^2}\bigg)\,,
\end{align}
where in the massless limit, $m_q^2\to0$,
\begin{align}
 \int_0^1\mathrm dz(1-z)z\ln\frac1{1-z(1-z)q^2/m_q^2}\to&\frac5{18}-\frac16\ln\frac{q^2}{m_q^2}+i\frac{\pi}6\,,\nonumber\\
 \int_0^1\mathrm dz(1-z)z\ln^2\frac1{1-z(1-z)q^2/m_q^2}\to&\frac{28}{27}+i\frac{5\pi}9-\frac{2\pi^2}9-\frac59\ln\frac{q^2}{m_q^2}-i\frac{\pi}3\ln\frac{q^2}{m_q^2}\nonumber\\
 &+\frac16\ln^2\frac{q^2}{m_q^2}\,,
\end{align}
and the residual $m_q^2$ dependence cancels to $\mathcal O(\epsilon)$ in eq.~(\ref{eq:l1}).
With this, the matrix elements are evaluated to
\begin{align}\label{eq:PE1l}
 \langle l\ell\ell|P_1|h\rangle_{1\text{loop}}&=-\frac89q_i\,I(m_q^2)\frac{\alpha_s}{4\pi}\langle P_9\rangle_\text{tree}\,,\nonumber\\
 \langle l\ell\ell|P_2|h\rangle_{1\text{loop}}&=\frac34\langle l\ell\ell|P_1|h\rangle_{1\text{loop}}\,,\nonumber\\
 \langle l\ell\ell|P_4|h\rangle_{1\text{loop}}&=-\frac43q_e\left(\frac{\mu^2}{m_c^2}\right)^\epsilon\left(-1-\epsilon\int_0^1\mathrm dz\ln\frac1{1-z(1-z)q^2/m_c^2}\right)\frac{\alpha_s}{4\pi}\langle P_7\rangle_\text{tree}\nonumber\\
 &-\frac89q_e\left(I(m_c^2)+I(0)\right)\frac{\alpha_s}{4\pi}\langle P_9\rangle_\text{tree}\,,\nonumber\\
 \langle l\ell\ell|E_{11}|h\rangle_{1\text{loop}}&=-4\left(-\frac89q_i\right)\left(\frac{\mu^2}{m_q^2}\right)^\epsilon\int_0^1\mathrm dz(1-z)z\left(-6-6\epsilon\ln\frac1{1-z(1-z)q^2/m_q^2}\right)\nonumber\\
 &\times\frac{\alpha_s}{4\pi}\langle P_9\rangle_\text{tree}\,,\nonumber\\
 \langle l\ell\ell|E_{12}|h\rangle_{1\text{loop}}&=\frac34\langle l\ell\ell|E_{11}|h\rangle_{1\text{loop}}\,.
\end{align}
We have checked that expanding the matrix elements in small $q^2$ and in the limit $m_q^2\to0$ yields the results in \cite{Asatryan:2001zw} and \cite{Seidel:2004jh}, respectively.

Following \cite{Asatryan:2001zw}, the renormalization of the mass $m_q$ is given by the replacement $m_q\to Z_{m_q}m_q$ in $\langle l\ell\ell|P_{1,2}|h\rangle_{1\text{loop}}$ of eq.~(\ref{eq:PE1l}).
The mass renormalization constants $Z_m$ in the \gls{msbar} and the pole mass scheme are \cite{Bobeth:1999mk}
\begin{align}
 &Z_m^{\overline{\text{MS}}}=1+\frac{\alpha_s}{4\pi}\left(-\frac4{\epsilon}\right)+\mathcal O(\alpha_s^2)\,,\nonumber\\
 &Z_m^{\text{pole}}=1+\frac{\alpha_s}{4\pi}\left(-\frac4{\epsilon}-4\ln\frac{\mu^2}{m^2}-\frac{16}3\right)+\mathcal O(\alpha_s^2)\,.
\end{align}
Expanding the matrix elements at $\mathcal O(\tfrac{\alpha_s}{4\pi})$ and $\mathcal O(\epsilon^0)$ gives the counterterms $F_{i,m_q}^{\text{ct}(9)}$ and $F_{i,m_q}^{\text{ct}(7)}=0$.
We have checked for both schemes that expanding the counterterm $F_{i,m_q}^{(9)}$ in small $q^2$ yields the results in \cite{Asatryan:2001zw}.

{\it Factorizable} form factors represented by class 5 in figure~\ref{fig:diagrams1to5} are found to be renormalized separately by the mass renormalization and mixing into evanescent operators, i.e.\ $a^{11}_{2\,11}=\tfrac12$ in eq.~(\ref{eq:a11}), and $a_{29}^{12}=\tfrac49q_i$ in eq.~(\ref{eq:a12}) are the only non-zero coefficients for the counterterms of $F_2^{(9fac)}$; recall that the $\langle P_1\rangle$ induced factorized form factor is given by color factors and the dipole factorized form factors are zero.

Finally, the renormalized form factors are given by subtracting $F_i^{\text{ct}(7,9)}=(F_{i\to7,9}^{\text{ct}(7,9)}+F_{i\to4\text{quark}}^{\text{ct}(7,9)}+F_{i,m_q}^{\text{ct}(7,9)})$ from the unrenormalized form factors.
Note that wave function renormalization of the external quark fields would need to be taken into account only if we wanted to include the diagrams of figure~\ref{fig:diagram6} \cite{Asatryan:2001zw}.
We have checked that $F_i^{\text{ct}(7)}|_{\epsilon^0}$ agrees with the results in \cite{Ghinculov:2003qd}.

\subsection{Numerical evaluation}
\label{sec:numerical_evaluation}

The analytical results for the renormalized form factors as found in the previous section are provided as supplemented files, see appendix \ref{app:supplemented_files}.
In this section, we describe the numerical evaluation of the \glspl{mi} \cite{Bell:2014zya}, which are expressed in terms of \glspl{hpl}.
Note that the undetermined ${\tilde M_{19}{}}_{\big|\epsilon^4}$ \cite{Bell:2014zya} in the \glspl{mi} cancels in the form factors.
In accordance with \cite{Bell:2014zya}, we choose the analytical continuation by subtracting from the internal mass an infinitesimal imaginary part $\eta>0$, i.e.\ $m_i^2\to m_i^2-i\eta$.
For negative invariant momentum transfer, an infinitesimal imaginary part is subtracted from $q^2$, and $m_i^2\to m_i^2+i\eta$.
Note that for positive invariant momentum transfer the addition of an infinitesimal imaginary part to $q^2$ is not necessary, i.e.\ real $q^2$ can be chosen.
For the evaluation of the factorizable results, we find that real $q^2$ are sufficient for any momentum transfer.  
We write the \glspl{hpl} as \glspl{gpl}, see, e.g., \cite{Bell:2014zya} (and references therein) and feed them into the computer package \texttt{lieevaluate} \cite{Frellesvig:2016ske}.
Other packages for the numerical evaluation of \glspl{gpl} can be found in \cite{Vollinga:2004sn,Kirchner:2016hmb}.

While evaluating the expressions with \texttt{lieevaluate}, we encounter the undefined function $\theta(0)$ and the function \texttt{MyP} \cite{Frellesvig:2016ske} that is divergent for some arguments of \glspl{gpl} \cite{Frellesvig:2016ske}.
We regulate the former by a perturbation of these arguments, see \cite{Frellesvig:2016ske}.
We have numerically checked that the form factors are insensitive to the actual choice of such perturbations.
The divergences due to the function \texttt{MyP} cancel in the form factors within the numerical precision.

We have checked that the numerical evaluation of the \glspl{mi} with \texttt{lieevaluate} yields a precision better than $\num{e-6}$ with respect to \cite{Bell:2014zya}.\
\footnote{We acknowledge the authors of \cite{Bell:2014zya} for providing additional code on their work.}
Moreover, numerical agreement of the \glspl{mi} in terms of \glspl{hpl} is found by use of the package \texttt{HypExp 2} \cite{Maitre:2007kp,Huber:2007dx} and also via numerical integration within \texttt{Mathematica}, yet the convergence is partially slow.

The analytical expressions are lengthy and their numerical evaluation is involved.
Hence, we evaluate the form factors for fixed mass parameters and at different $q^2$ points.
Note that the numerical evaluation close to the kinematical endpoints $q^2=0,m_h^2$ is sensitive to the ratio of mass parameters and $q^2$.
Finally, we fit the points.

Subtracting the counterterm form factors from the unrenormalized form factors, the $\tfrac1{\epsilon^2}$ and $\tfrac1{\epsilon}$ divergences cancel numerically.
We have checked our calculation against the ones of \cite{Asatryan:2001zw,Seidel:2004jh,Greub:2008cy} for \mbox{$b\to(d,s)$} transitions, finding numerical agreement for different $q^2$, scales, mass schemes, and parameters, see the next section.
For \mbox{$c\to u\gamma$} transitions, the effective dipole Wilson coefficient at $q^2=0$ induced by $\langle P_2\rangle$ was calculated in \cite{Greub:1996wn}.
Adding the constant terms given in \cite{deBoer:2015boa} we have checked the calculations, finding numerical agreement.

\section{Results}
\label{sec:results}

\begin{figure}
 \centering
 \includegraphics[width=0.475\textwidth]{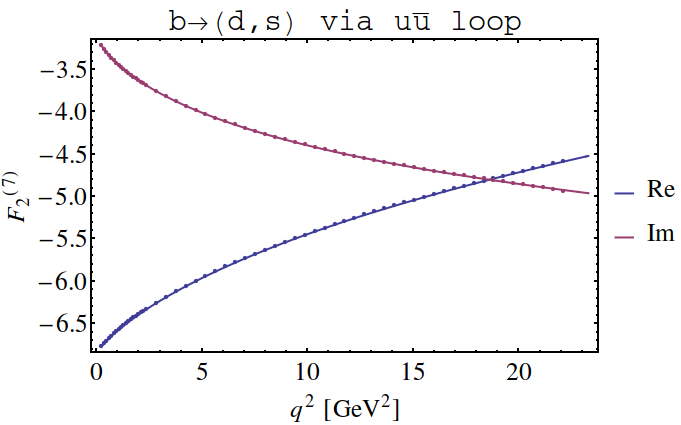}
 \hspace{0.5em}
 \includegraphics[width=0.475\textwidth]{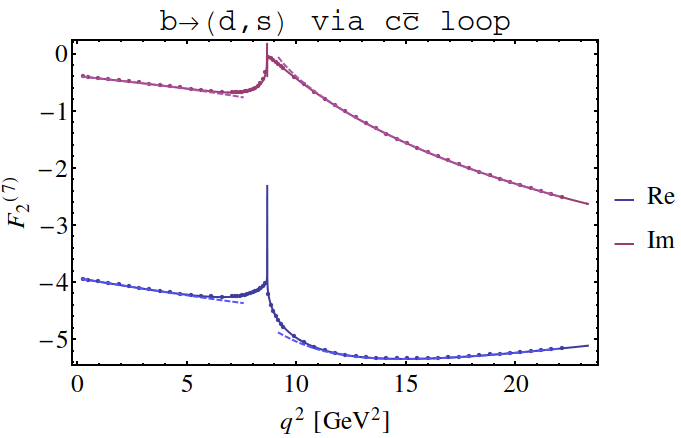}
 \\[1ex]
 \includegraphics[width=0.425\textwidth]{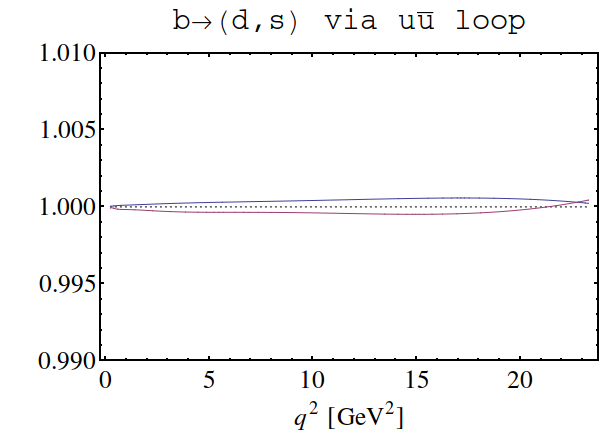}
 \hspace{2.5em}
 \includegraphics[width=0.425\textwidth]{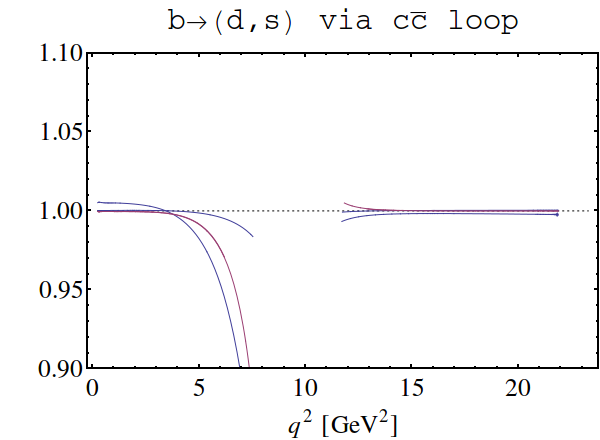}
 \hspace{2em}
 \caption{Real (blue) and imaginary (purple) parts of the form factor $F_2^{(7)}$ induced by $\langle P_2^{(u)}\rangle$ (left) and $\langle P_2^{(c)}\rangle$ (right) for \mbox{$b\to(d,s)$} transitions.
 The solid lines are fitted to the points, which represent the results of the numerical evaluation.
 The dashed lines show the expansions of \cite{Asatryan:2001zw,Greub:2008cy}, whereas, in the upper left plot, the result of \cite{Seidel:2004jh} agrees with the solid lines.
 The lower plots show the ratio of the fitted curves with respect to results of \cite{Seidel:2004jh} (lower left) and the expansions of \cite{Asatryan:2001zw,Greub:2008cy} (lower right), where $\mu=\tfrac12m_b,2m_b$.
 A ratio of one is indicated by the black dotted line in the lower plots.}
 \label{fig:results_F27b}
\end{figure}

\begin{figure}
 \centering
 \includegraphics[width=0.475\textwidth]{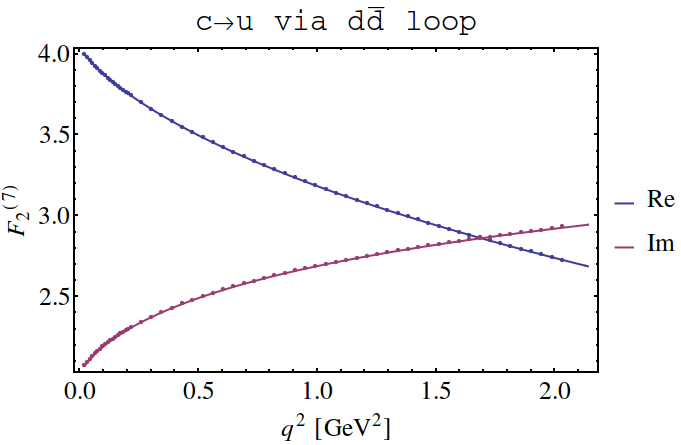}
 \hspace{0.5em}
 \includegraphics[width=0.475\textwidth]{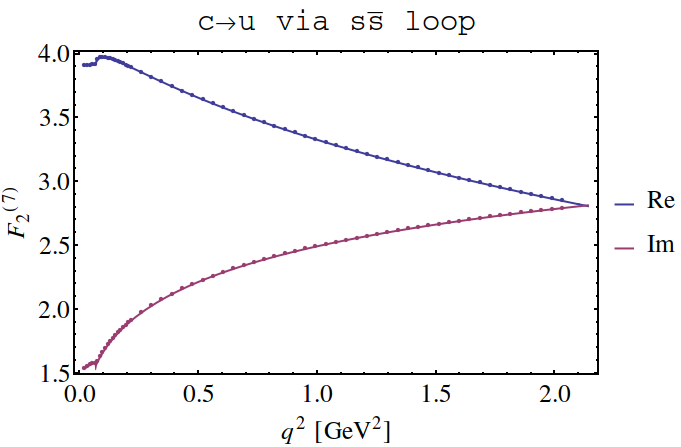}
 \caption{Real (blue) and imaginary (purple) parts of the form factor $F_2^{(7)}$ induced by $\langle P_2^{(d)}\rangle$ (left) and $\langle P_2^{(s)}\rangle$ (right) for \mbox{$c\to u$} transitions.
 The solid lines are fitted to the points, which represent the results of the numerical evaluation.}
 \label{fig:results_F27c}
\end{figure}

\begin{figure}
 \centering
 \includegraphics[width=0.475\textwidth]{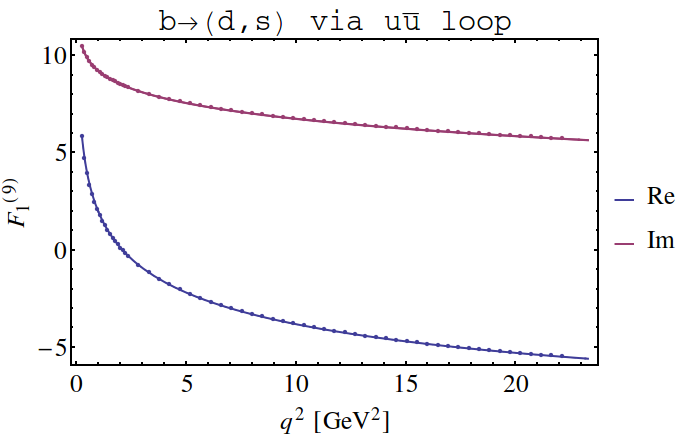}
 \hspace{0.5em}
 \includegraphics[width=0.475\textwidth]{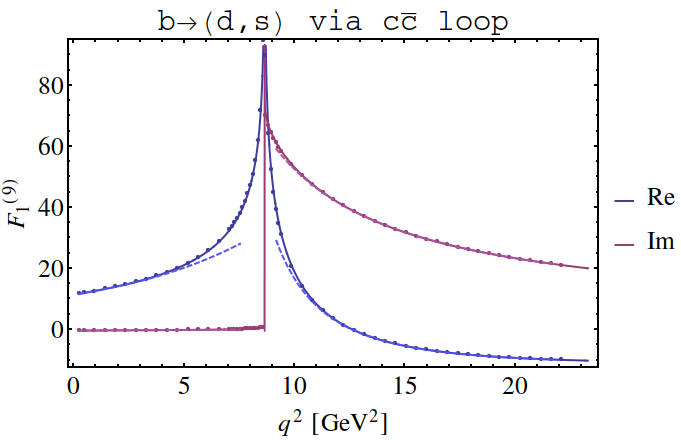}
 \\[1ex]
 \includegraphics[width=0.475\textwidth]{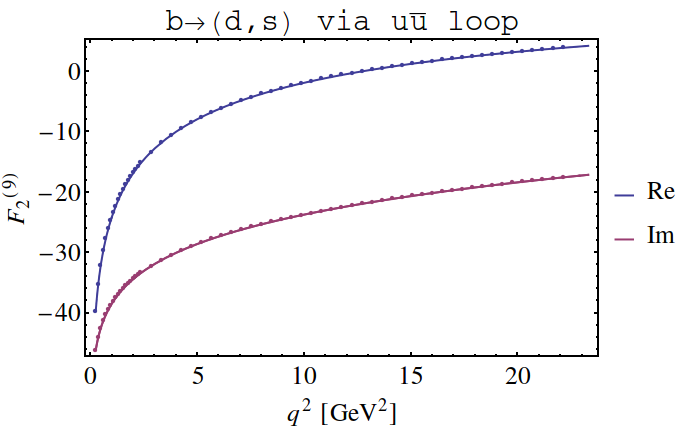}
 \hspace{0.5em}
 \includegraphics[width=0.475\textwidth]{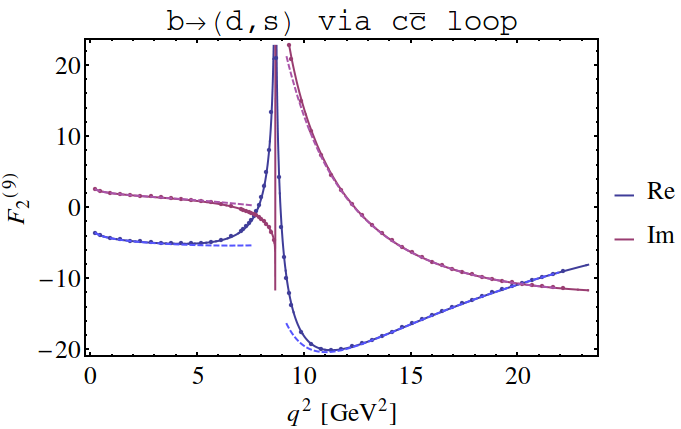}
 \caption{Real (blue) and imaginary (purple) parts of the form factor $F_{1,2}^{(9)}$ induced by $\langle P_1^{(u)}\rangle$ (upper left), $\langle P_1^{(c)}\rangle$ (upper right), $\langle P_2^{(u)}\rangle$ (lower left), and $\langle P_2^{(c)}\rangle$ (lower right) for \mbox{$b\to(d,s)$} transitions.
 The solid lines are fitted to the points, which represent the results of the numerical evaluation.
 The dashed lines show the expansions of \cite{Asatryan:2001zw,Greub:2008cy}, whereas, in the left plots, the results of \cite{Seidel:2004jh} agree with the solid lines.}
 \label{fig:results_F1929b}
\end{figure}

\begin{figure}
 \centering
 \includegraphics[width=0.475\textwidth]{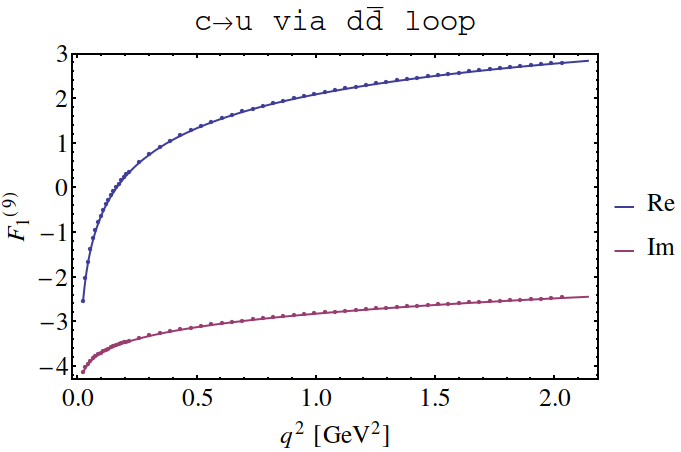}
 \hspace{0.5em}
 \includegraphics[width=0.475\textwidth]{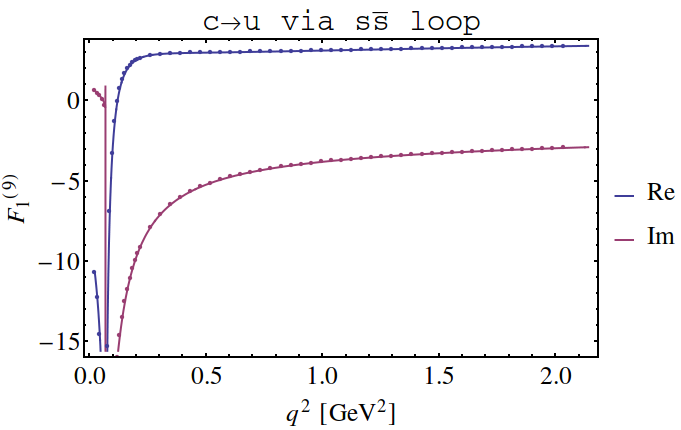}
 \\[1ex]
 \includegraphics[width=0.475\textwidth]{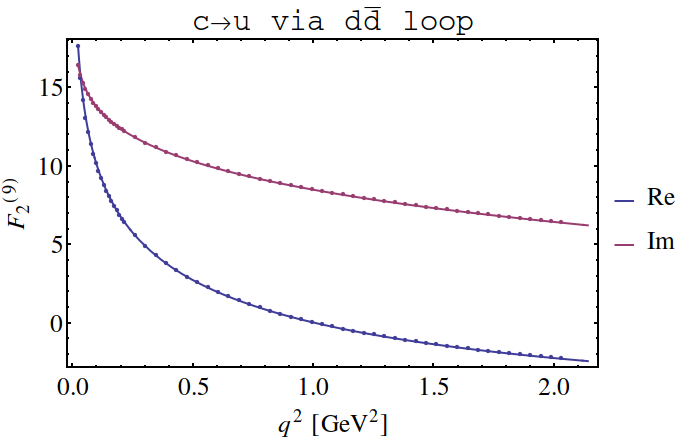}
 \hspace{0.5em}
 \includegraphics[width=0.475\textwidth]{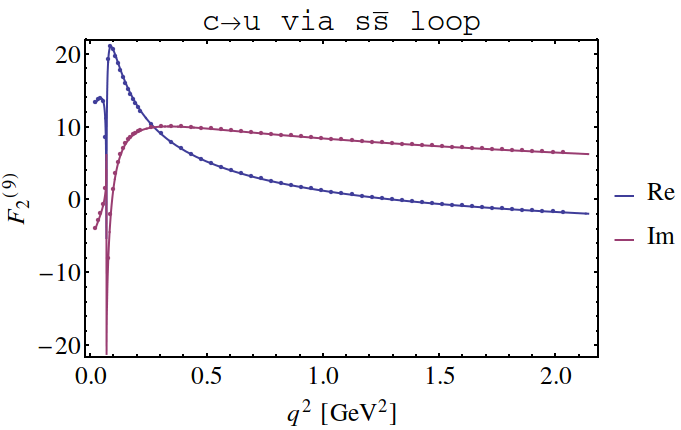}
 \caption{Real (blue) and imaginary (purple) parts of the form factor $F_{1,2}^{(9)}$ induced by $\langle P_1^{(d)}\rangle$ (upper left), $\langle P_1^{(s)}\rangle$ (upper right), $\langle P_2^{(d)}\rangle$ (lower left), and $\langle P_2^{(s)}\rangle$ (lower right) for \mbox{$c\to u$} transitions.
 The solid lines are fitted to the points, which represent the results of the numerical evaluation.}
 \label{fig:results_F1929c}
\end{figure}

Our fitted results of the renormalized form factors $F_{1,2}^{(7,9)}$ induced by $\langle l\ell\ell|P_{1,2}|h\rangle$ for \mbox{$b\to(d,s)$} and \mbox{$c\to u$} transitions are shown in figures~\ref{fig:results_F27b}-\ref{fig:results_F1929c} for $q^2\in[0,m_h^2]$.
For comparison, we also show the results of \cite{Asatryan:2001zw,Seidel:2004jh,Greub:2008cy}.
We use the (pole) masses
\begin{align}\label{eq:masses}
 &m_b=\SI{4.85}{GeV}\,,&&m_c=\SI{1.47}{GeV}\,,&&m_s=\SI{0.13}{GeV}\,,
\end{align}
$m_{u,d}\approx0$ and set $\mu=m_h$ if not stated otherwise.

We note the following:
\begin{itemize}
 \item The results of \cite{Asatryan:2001zw,Seidel:2004jh,Greub:2008cy} agree well with our results, hence the former are only partially visible in the plots.
 \item The form factors are divergent at the internal quark pair mass squared $q^2=(2m_i)^2$.
 \item The lower left plot of figure~\ref{fig:results_F27b} indicates a numerical precision of our results better than $\num{e-3}$ with respect to the analytical result of \cite{Seidel:2004jh}.
 \item The lower right plot of figure~\ref{fig:results_F27b} shows agreement with the expansion of \cite{Greub:2008cy} at high $q^2$.
 Note that we do not plot the ratio close to the kinematical endpoint $q^2=m_b^2$, where the evaluation of the result of \cite{Greub:2008cy} yields oscillations.
 At low $q^2$, a breakdown of the convergence of the expansion of \cite{Asatryan:2001zw} is visible around $q^2\sim\SI{5}{GeV}^2$, where the sensitivity on $\mu$ increases.
 Similar conclusions are drawn from figure~\ref{fig:results_F1929b}.
\end{itemize}
Furthermore, note that our definition of the form factors and the one in \cite{Asatryan:2001zw,Greub:2008cy} differs by a minus sign.
Recall that $F_1^{(7)}=\tfrac{-1}6F_2^{(7)}$.
For $F_{1,2}^{(9)}$, the numerical precision of our results is better than $\num{e-2}$ with respect to the results of \cite{Asatryan:2001zw,Seidel:2004jh,Greub:2008cy} for \mbox{$b\to(d,s)$} transitions.

We deduce that the overall numerical precision of our result is better than a percent.
To have ready-to-use result at hand, we provide our results for the masses of eq.~(\ref{eq:masses}) as supplemented files, see appendix~\ref{app:supplemented_files}.

\begin{figure}
 \centering
 \includegraphics[width=0.425\textwidth]{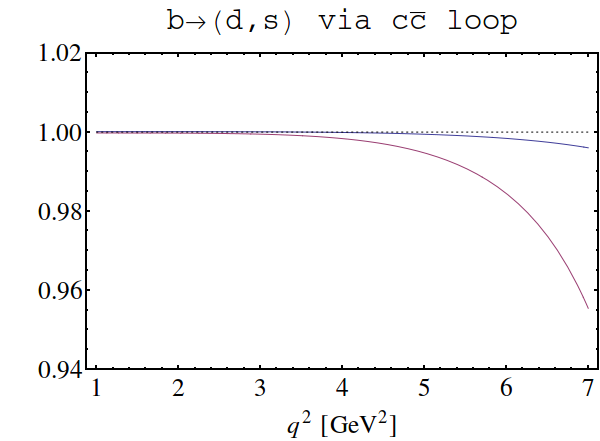}
 \hspace{2.5em}
 \includegraphics[width=0.425\textwidth]{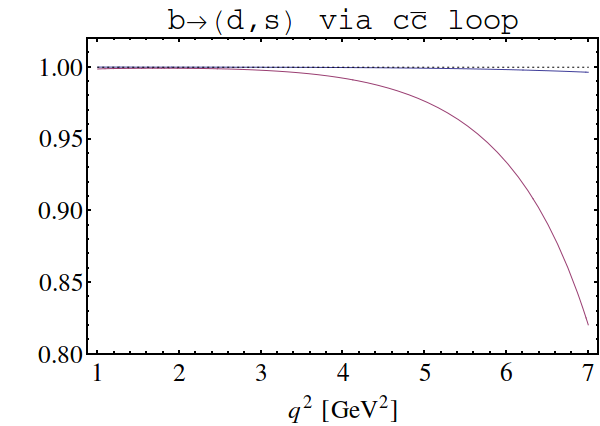}
 \hspace{2em}
 \caption{Ratios $\tfrac{\mathrm{Re}[\tilde C^\text{eff}]^\text{this}}{\mathrm{Re}[\tilde C^\text{eff}]^\text{\cite{Asatryan:2001zw}}}$ (blue) and $\tfrac{\mathrm{Im}[\tilde C^\text{eff}]^\text{this}}{\mathrm{Im}[\tilde C^\text{eff}]^\text{\cite{Asatryan:2001zw}}}$ (purple) of the effective Wilson coefficients $\tilde C_7^\text{eff}$ (left) and $\tilde C_9^\text{eff}$ (right) for \mbox{$b\to(d,s)$} transitions induced by a massive internal charm quark.
 We follow \cite{Ghinculov:2003qd,Greub:2008cy}, adding the results of this paper and the expansion of \cite{Asatryan:2001zw} for $\mu=\SI{5}{GeV}$.
 A ratio of one is indicated by the black dotted line.}
 \label{fig:lowq2ratio_tildeC79eff}
\end{figure}

\begin{figure}
 \centering
 \includegraphics[width=0.475\textwidth]{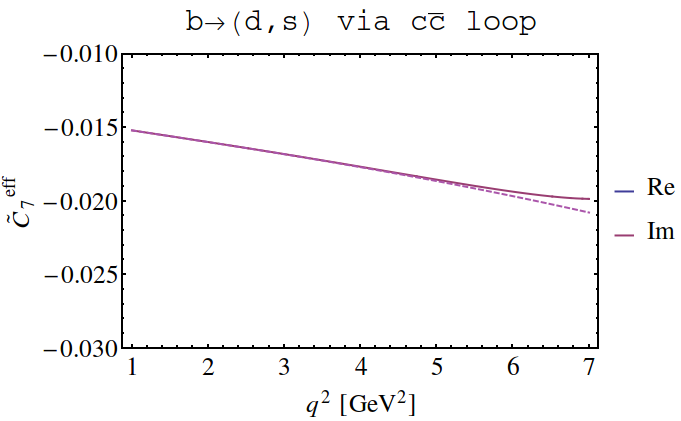}
 \hspace{0.5em}
 \includegraphics[width=0.475\textwidth]{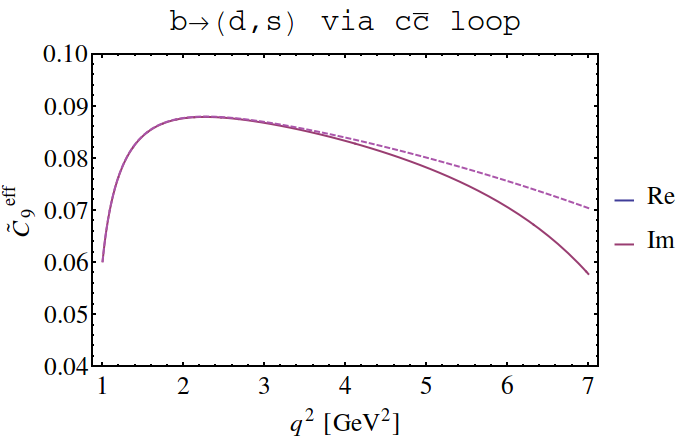}
 \caption{Imaginary parts of the effective Wilson coefficients $\tilde C_7^\text{eff}$ (left) and $\tilde C_9^\text{eff}$ (right) for \mbox{$b\to(d,s)$} transitions induced by a massive internal charm quark.
 We follow \cite{Ghinculov:2003qd,Greub:2008cy}, adding the results of this paper (solid) and the expansion of \cite{Asatryan:2001zw} (dashed) for $\mu=\SI{5}{GeV}$.}
 \label{fig:lowq2Im}
\end{figure}

Next, we comment on the impact of the new results for \mbox{$b\to(d,s)$} transitions.
For massless internal quarks an independent analytical result is provided in \cite{Seidel:2004jh}.
The high $q^2$ range for massive internal quarks is well approximated by the results of \cite{Greub:2008cy}.
On the other side, compared to the low $q^2$ range \cite{Asatryan:2001zw} for massive internal quarks, our results indicate significant corrections around $q^2\sim\SI{5}{GeV}^2$.
Following \cite{Ghinculov:2003qd} and \cite{Greub:2008cy} for the matrix element of the chromomagnetic operator, the effective Wilson coefficients $\tilde C_{7,9}^\text{eff}$ at \gls{nnll} order are shown in figures~\ref{fig:lowq2ratio_tildeC79eff},~\ref{fig:lowq2Im}.
We add our results and, for comparison, the results of \cite{Asatryan:2001zw} for $\SI{1}{GeV}^2\le q^2\le\SI{7}{GeV}^2$.
Note that for phenomenological purposes $\tilde P_{7,9}=\tfrac{\alpha_s}{4\pi}P_{7,9}$, thus $\tilde C_{7,9}=\tfrac{4\pi}{\alpha_s}C_{7,9}$.

One observes that the real parts are stable, whereas corrections to imaginary parts increase to several percent towards higher $q^2$.
The effective Wilson coefficients obey the hierarchies $\mathrm{Re}[\tilde C_{7,9}^\text{eff}]\gg\mathrm{Im}[\tilde C_{7,9}^\text{eff}]$.
Thus, observables which only depend on the real parts or magnitudes of the effective Wilson coefficients are marginally affected by our results as long as the results of \cite{Asatryan:2001zw,Seidel:2004jh,Greub:2008cy} are taken into account.
On the other hand, effects on observables which depend on the imaginary parts of the effective Wilson coefficients are significant in the low $q^2$ range.

\begin{figure}
 \centering
 \includegraphics[width=0.475\textwidth]{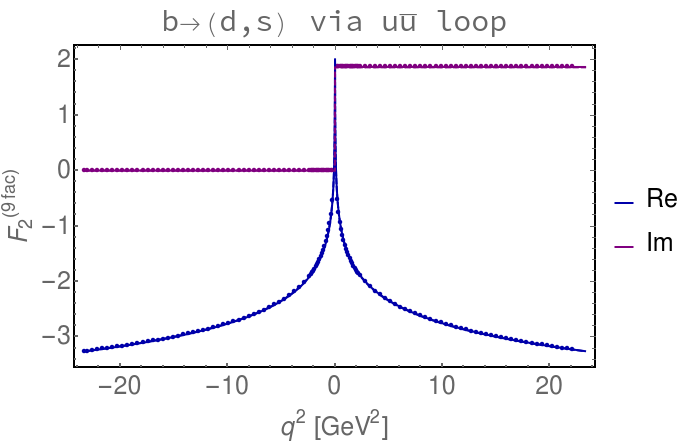}
 \hspace{0.5em}
 \includegraphics[width=0.475\textwidth]{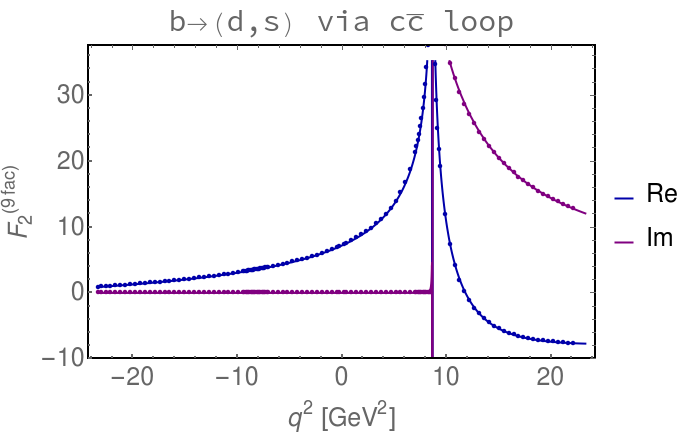}
 \\[1ex]
 \includegraphics[width=0.475\textwidth]{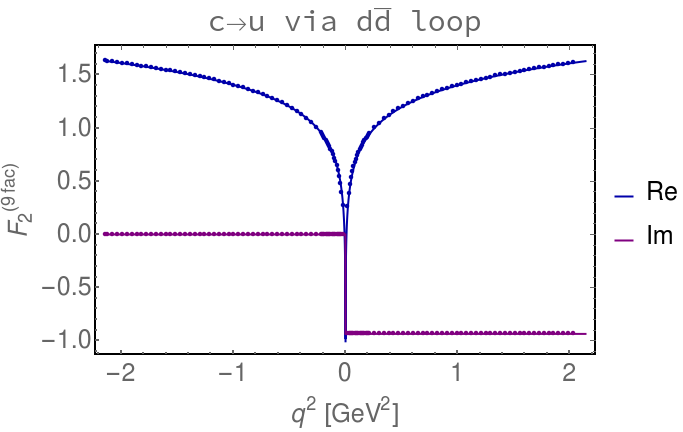}
 \hspace{0.5em}
 \includegraphics[width=0.475\textwidth]{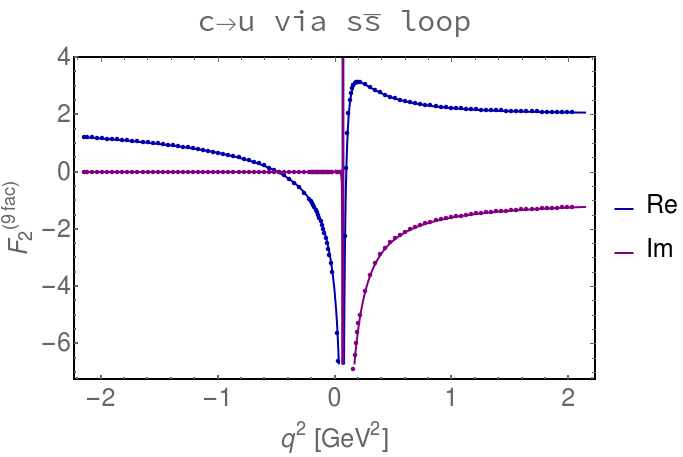}
 \caption{Real (blue) and imaginary (purple) parts of the form factor $F_2^{(9fac)}$ induced by $\langle P_2^{(u)}\rangle$ (upper left), and $\langle P_2^{(c)}\rangle$ (upper right) for \mbox{$b\to(d,s)$} transitions and $\langle P_2^{(d)}\rangle$ (lower left), and $\langle P_2^{(s)}\rangle$ (lower right) for \mbox{$c\to u$} transitions.
 The solid lines are fitted to the points, which represent the results of the numerical evaluation.
 In the lower left plot, the results (three times the function \texttt{C}) of \cite{Seidel:2004jh} agree with the solid lines.}
 \label{fig:results_F29Ebc}
\end{figure}

Our fitted results of the renormalized form factors $F_2^{(9fac)}$, class 5 in figure~\ref{fig:diagrams1to5}, are shown in figure~\ref{fig:results_F29Ebc} for $q^2\in[-m_h^2,m_h^2]$.

We note that the factorized form factors are smooth below the internal quark pair threshold and real in the Euclidean region.
One observes that for vanishing internal masses, see left plots in figure~\ref{fig:results_F29Ebc}, the shapes only differ by the ratio of internal charges and $q^2$ scales with the ratio of external masses.

\section{Summary}
\label{sec:summary}

In this paper, we calculated the effective Wilson coefficients for heavy to light \gls{fcnc} transitions induced by the matrix elements of current-current operators at two loop.
The results are valid for arbitrary momentum transfer and masses, when the light mass is neglected.
For the \glspl{mi}, we used the works of \cite{Bell:2014zya,Frellesvig:2016ske}.
Our computation extends previous works on \mbox{$b\to(d,s)\ell^+\ell^-$} transitions \cite{Asatryan:2001zw,Asatrian:2003vq,Seidel:2004jh,Greub:2008cy} and is new for \mbox{$c\to u\ell^+\ell^-$} transitions.
We found significant corrections to the imaginary parts of the effective Wilson coefficients for \mbox{$b\to(d,s)$} transitions in the large hadronic recoil range, see figure~\ref{fig:lowq2Im}, which should be included in future analyses, including \gls{bsm} studies.
Other corrections for \mbox{$b\to(d,s)$} transitions were found to be marginal.
For \mbox{$c\to u$} transitions and an application to leptoquark models, we refer to \cite{deBoer}.
Along with this paper we provide supplemented files, containing our analytical results and fitted results for a specific set of parameters.
Finally, our calculation is an independent check of the results of \cite{Bell:2014zya,Frellesvig:2016ske} and \cite{Asatryan:2001zw,Asatrian:2003vq,Seidel:2004jh,Greub:2008cy}, with which we agree in the corresponding limits.

\section*{Acknowledgements}

We acknowledge Tobias Huber for several useful discussions.
We thank Christoph Bobeth and Gudrun Hiller for useful comments on the manuscript.
This work has been supported in part by the DFG Research Unit FOR 1873 ``Quark Flavour Physics and Effective Field Theories''.

\begin{appendix}

\section{Supplemented files}
\label{app:supplemented_files}

The analytical results and the results fitted to points covering the range $|q^2|\le m_h^2$ for the masses of eq.~(\ref{eq:masses}) of the renormalized form factors induced by $\langle l\ell\ell|P_{1,2}|h\rangle$ are supplemented to the source files of this paper at \url{https://arxiv.org/abs/1707.00988}.
These supplemented files are described in this appendix and can be used with, e.g., \texttt{Mathematica}.
Recall that $F_1^{(7)}=\tfrac{-1}6F_2^{(7)}$, $F_1^{(9fac)}=\tfrac43F_2^{(9fac)}$, $F_{1,2}^{(7fac)}=0$ and that the light external mass is neglected.

The fitted results are provided by the files \texttt{fit\_F*.dat}, where the asterisk specifies the form factor and the transition, e.g.\ \texttt{fit\_F92\_btods\_ccbar.dat} denotes the form factor $F_2^{(9)}$ for \mbox{$b\to(d,s)$} transitions via a $c\bar c$ loop (induced by $\langle P_2^{(c)}\rangle$).
The fits are functions of polynomials and logarithms in $\texttt Q=q^2$ in units of $\text{GeV}^2$ and $\texttt{MU}=\mu^2/m_e^2$, where $m_e$ denotes the (heavy) external quark mass, e.g.\ $m_e=m_b$ for \mbox{$b\to(d,s)$} transitions.

The analytical results are provided by the files \texttt{F*.dat}, where the asterisk specifies the form factor and type of polylogarithm, e.g.\ \texttt{F92fac\_HPL.dat} denotes the form factor $F_2^{(9fac)}$ in terms of \glspl{hpl}.
The \texttt{HPL} files contain the most general and compact results of this paper, yet individual terms are literally divergent (the full expression is finite).
Converted to \glspl{gpl}, as provided by the \texttt{GPL} files, these divergences cancel.
Recall that a regularization may be necessary for the numerical evaluation, see section~\ref{sec:numerical_evaluation}.
Again, $\texttt Q=q^2$ and $\texttt{MU}=\mu^2/m_h^2$.
Furthermore, $\texttt{qe}=q_e$ and $\texttt{qi}=q_i$ are the external and internal charges, respectively, $\texttt{me}=m_e$ and $\texttt{mi}=m_i$ are the external and internal masses.
The mass scheme parameter $\texttt{massscheme}\to0$ for \gls{msbar} masses and $\texttt{massscheme}\to(-4\ln\tfrac{\mu^2}{m_i^2}-\tfrac{16}3)$ in the pole mass scheme.
The arguments of the \glspl{hpl} are defined by the weight functions as \cite{Bell:2014zya}
\begin{align}
 &f_{w^+}(x)=f_w(x)+f_{-w}(x)\,,&&f_{w^-}(x)=f_w(x)-f_{-w}(x)
\end{align}
with
\begin{align}
 &w_1=1\,,&&w_2=r\,,&&w_3=\frac{r^2+1}2\,,&&w_4=1+\sqrt{1-r^2}\,,&&w_5=1-\sqrt{1-r^2}\,.
\end{align}
Furthermore,
\begin{align}
 &r=\sqrt{1-4z_i}\,,&&z_i=\frac{m_i^2}{m_e^2}
\end{align}
and
\begin{align}
 &s=\sqrt{1-\frac{4z_i}{\bar u}}\,,&&s_1=\sqrt{1-\frac4{\bar u}}\,,&&\bar u=\frac{q^2}{m_e^2}\,,\nonumber\\
 &t=\frac{1-s_1}2+\frac{1+s_1}2\sqrt{1+\frac{2(1-r^2)(1-s_1)}{(1+s_1)^2}}\,,&&t_0=e^{i\pi/3}r+e^{-i\pi/3}\,,\nonumber\\
 &v=\frac{1+s_1}2+\frac{1-s_1}2\sqrt{1+\frac{2(1-r^2)(1+s_1)}{(1-s_1)^2}}\,,&&v_0=e^{-i\pi/3}r+e^{i\pi/3}\,.
\end{align}
Finally, $\texttt{I}=i$, $\texttt{Pi}=\pi$, $\texttt{EulerGamma}=\gamma_E$, $\texttt{Zeta[s]}=\zeta(s)$, $\texttt{Sqrt[z]}=\sqrt{z}$, $\texttt{Log[z]}=\ln(z)$ and $\texttt{PolyLog[n,z]}=Li_n(z)$, following the \texttt{Mathematica} notation.

\end{appendix}

\printbibliography

\end{document}

%% file: glossary.tex
\newacronym{adm}{ADM}{anomalous dimension matrix}
\newacronym{bsm}{BSM}{beyond the standard model}
\newacronym{ckm}{CKM}{Cabibbo--Kobayashi--Maskawa}
\newacronym{fcnc}{FCNC}{flavor changing neutral current}
\newacronym{gim}{GIM}{Glashow--Iliopolus--Maiani}
\newacronym{gpl}{GPL}{generalized/Goncharov polylogarithm}
\newacronym{hpl}{HPL}{harmonic polylogarithm}
\newacronym{ibp}{IBP}{integration by parts}
\newacronym{li}{LI}{Lorentz invariance}
\newacronym{lo}{LO}{leading order}
\newacronym{mi}{MI}{master integral}
\newacronym{msbar}{$\overline{\text{MS}}$}{modified minimal subtraction}
\newacronym{nnll}{NNLL}{next-to-next-to leading logarithmic}
\newacronym{nlo}{NLO}{next-to leading order}
\newacronym{ope}{OPE}{operator product expansion}
\newacronym{qcd}{QCD}{quantum chromodynamics}
\newacronym{sm}{SM}{standard model}